\documentclass[10pt,preprint]{aastex}
\begin{document}

\title{Discovery of very extended emission-line region around the Seyfert 2 
galaxy NGC 4388
\footnote{Based on data collected at Subaru Telescope, which is operated by the
          National Astronomical Observatory of Japan.}}

\author{
Michitoshi Yoshida\altaffilmark{2,3},  
Masafumi Yagi\altaffilmark{2}, 
Sadanori Okamura\altaffilmark{4,5},
Kentaro Aoki\altaffilmark{6,7}, 
Youichi Ohyama\altaffilmark{8},
Yutaka Komiyama\altaffilmark{8},
Naoki Yasuda\altaffilmark{6},
Masanori Iye\altaffilmark{2},
Nobunari Kashikawa\altaffilmark{2},
Mamoru Doi\altaffilmark{7},
Hisanori Furusawa\altaffilmark{4},
Masaru Hamabe\altaffilmark{11}, 
Masahiko Kimura\altaffilmark{10}, 
Masayuki Miyazaki\altaffilmark{4},
Satoshi Miyazaki\altaffilmark{9},
Fumiaki Nakata\altaffilmark{4},
Masami Ouchi\altaffilmark{4},
Maki Sekiguchi\altaffilmark{10},
Kazuhiro Shimasaku\altaffilmark{4},
AND
Hiroshi Ohtani\altaffilmark{12}
}

\vspace {3cm}

\altaffiltext{2}{Optical and Infrared Astronomy Division, National Astronomical Observatory, 
Mitaka, Tokyo, 181-8588 Japan.}

\altaffiltext{3}{Okayama Astrophysical Observatory, National Astronomical Observatory,
Kamogata, Okayama, 719-0232 Japan; yoshida@oao.nao.ac.jp.}

\altaffiltext{4}{Department of Astronomy, School of Science, University of Tokyo,
Tokyo, 113-0033 Japan.}

\altaffiltext{5}{Research Center for the Early Universe, School of Science,
University of Tokyo, Tokyo, 113-0033 Japan.}

\altaffiltext{6}{Astronomical Data Analysis Center, National Astronomical Observatory, Mitaka,
Tokyo, 181-8588 Japan.}

\altaffiltext{7}{Institute of Astronomy, School of Science, University of Tokyo, Mitaka, 
Tokyo, 181-8588 Japan.}

\altaffiltext{8}{Subaru Telescope, National Astronomical Observatory of Japan, 650 North
A'Ohoku Place, Hilo, HI 96720 USA.}

\altaffiltext{9}{Advanced Technology Center, National Astronomical Observatory, Mitaka, 
Tokyo, 181-8588 Japan.}

\altaffiltext{10}{Institute for Cosmic Ray Research, University of Tokyo, Kashiwa, Chiba, 
277-8582 Japan.}

\altaffiltext{11}{Department of Mathematical and Physical Sciences, Faculty of Science,
Japan Women's University, Bunkyo-ku, Tokyo, 112-8681 Japan.}

\altaffiltext{12}{Department of Astronomy, Faculty of Science,
Kyoto University, Sakyo-ku, Kyoto, 606-8502 Japan.}


\begin{abstract}

  We found a very large, $\sim 35$ kpc, emission-line region
around the Seyfert
type 2 galaxy NGC 4388, using deep narrow-band imaging with the prime
focus camera (Suprime-Cam) of the Subaru telescope.  This region
consists of many faint gas clouds or filaments, and extends
northeastwards from the galaxy.  The typical H$\alpha$ luminosity
$L$(H$\alpha$) of the clouds is $\sim 10^{37}$ erg s$^{-1}$, and the
total $L$(H$\alpha$) of the region within 10 kpc from the nucleus is
$\sim 2 \times 10^{38}$ erg s$^{-1}$, which corresponds to an ionized
gas mass of $\sim 10^{5} M_{\sun}$.  The map of the emission-line
intensity ratio $I$([\ion{O}{3}])$/I$(H$\alpha$) indicates that the inner
($r<12$ kpc) region of the VEELR may be excited by nuclear ionizing
radiation.  The excitation mechanism of the outer ($r>12$ kpc) region
is unclear, but it is likely that the nuclear radiation is also a
dominant source of its ionization.  We discuss the origin of the
ionized gas.  Two plausible origins of the gas in the VEELR are
(i) the tidal debris resulting from a past interaction with a gas-rich
dwarf galaxy, i.e., a minor merger, or (ii) the interstellar medium of
NGC 4388, stripped by the ram pressure of the hot intracluster medium
of the Virgo cluster.

\end{abstract}




\section{INTRODUCTION}

  It is widely recognized that extended emission-line regions are
commonly associated with active galactic nuclei (AGNs), and that most
of them are photoionized by the nuclear power-law ionizing continuum.
The extended emission-line regions of AGNs are often in the shape of a
cone, whose apex corresponds to the nucleus.  This trend is generally
interpreted as evidence for anisotropy of the nuclear radiation of
AGNs (e.g., Pogge 1989; Mulchaey, Wilson, \& Tsvetanov 1996; Falcke,
Wilson \& Simpson 1998).

  NGC 4388 is one of the nearest Seyfert 2 galaxies, and the first
Seyfert to be found in the Virgo cluster (Phillips \& Malin 1982).
NGC 4388 is located within 1$^{\circ}$ of the two cD galaxies, M84
and M86, which lie at the center of the Virgo cluster
(Binggeli, Tamman \& Sandage 1987).  Although the recession velocity
of NGC 4388 (2540 km s$^{-1}$) is substantially larger than the
cluster mean velocity of 1100 km s$^{-1}$, its distance estimated from
the Tully-Fisher relation (16.7 Mpc) is close to that of the cluster
center (Yasuda, Okamura \& Fukugita 1997).  The spatial location of
the galaxy is thus very close to the core of the cluster.  At a
distance of 16.7 Mpc, 1\arcmin\ corresponds to 4.86 kpc.

  NGC 4388 is associated with a large, bright
optical emission-line region (Pogge 1988; Corbin, Baldwin \& Wilson
1988; Ayani \& Iye 1989; Veilleux et al. 1999a, hereafter VBCTM,
and references therein).  The extended emission-line region (EELR) of NGC
4388 is characterized by at least three components: 1) a low-ionization
disk component, 2) a high-ionization cone-shaped component
extending southwestwards from the nucleus (hereafter referred to as
the ``SW cone''), and 3) a large high-ionization plume extending northeastwards
from the nucleus up to 4 kpc above the plane of the galaxy
(hereafter referred to as the ``NE plume'') (Pogge 1988; Corbin et
al. 1988; Colina 1992; Falcke et al. 1998; VBCTM).  The low-ionization
component is attributed to a complex of disk \ion{H}{2} regions.
The excitation properties of the gas in the SW cone and the NE plume
resemble those of the nuclear narrow-line region gas, and are
consistent with power-law photoionization models, indicating that
anisotropic ionizing radiation from the Seyfert nucleus forms these
regions (Pogge 1988; Petitjean and Durret 1993).  The NE plume extends
further than the SW cone, which is in front of the galaxy disk and
consequently suffers less from dust obscuration.  The shape of the NE plume is
remarkable; it extends along P.A.$ \approx30^{\circ}$ from the
nucleus like a wide belt, and curls towards the northwest at
a distance of $\approx 3$ kpc (VBCTM). The total length of the NE plume
is up to 4 kpc.  The origin of the NE plume is still controversial,
although several models have been proposed: tidal
debris from a galaxy-galaxy encounter (Pogge 1988), gas stripped from the
galaxy disk by ram pressure induced by interaction with the intracluster
gas (Petitjean \& Durret 1993), ejecta from supernovae in the disk
(Corbin et al. 1988), or a nuclear outflow powered by the radio jet
(VBCTM).

  In this paper, we report the discovery of a very extended emission-line region
around NGC 4388, in deep, narrow-band images made with the Subaru
telescope.  The extension of the emission-line region is more than eight
times greater than that of the previously found emission,
extending as far as 35 kpc from the nucleus.  We describe the
observations and data reduction in $\S$2, and present the results in
$\S$3.  In $\S$4 we briefly discuss possible models for the origin of
the emission-line gas and comment on its relationship to the inner
emission-line region.  We summarize our study in $\S$5.

\section{OBSERVATIONS AND DATA REDUCTION}

  The narrow-band and broad-band imaging observations were made with
the Suprime-Cam (Miyazaki et al. 1998; Iye et al. 1997) attached at
the prime focus of the Subaru telescope (Kaifu et al. 2000) on 2001
March 24 and 2001 April 24 and 26.  In the March run, the camera
consisted of 9 CCDs; four SITe 2K $\times$ 4K CCDs and five MIT 2K
$\times$ 4K CCDs.  After the March run, all the SITe CCDs were
replaced by new MIT CCDs, so 10 MIT CCDs were used in the April
run.  The pixel size was 15 \micron, which corresponds to 0.2 arcsec on
the sky.  The H$\alpha$+[\ion{N}{2}] narrow-band imaging of NGC 4388
was performed on 2001 April 26 with a narrow-band filter whose central
wavelength, $\lambda_0$, and FWHM were $6600$ \AA\ and $100$ \AA,
respectively (filter N-A-L659).  The [\ion{O}{3}] $\lambda5007$
\AA\ imaging was carried out on 2001 April 24, using a filter whose
$\lambda_0$ and FWHM were $5020$ \AA\ and $100$ \AA\ respectively
(filter N-A-L503).  To subtract the continuum component from the
narrow-band images so as to extract a pure emission-line image, we
obtained $R_c$-band and $V$-band images, which allowed us to estimate
the continuum in the H$\alpha$+[\ion{N}{2}] image and [\ion{O}{3}]
image, respectively.  The broad-band data were taken on 2001 March 24.
The observations are summarized in Table 1.  The seeing was
0\farcs6 -- 0\farcs7 on the nights of the narrow-band imaging
observations, and 0\farcs9 -- 1\farcs0 on the nights of the
broad-band imaging.  The sky was clear, but not photometric, on these
nights; however, some photometric standard stars were obtained for
rough flux calibration.

  Data reduction was performed using IRAF and DASH (Mizumoto et al. 2000;
Yagi et al. 2000), which is a pipeline data reduction system in
a distributed computing environment. Basic data reduction procedures,
i.e., bias and dark subtraction, flat fielding, and distortion correction,
were carried out using a data reduction pipeline developed
specifically for Suprime-Cam data under the DASH environment.
Since significant non-uniformity ($\approx 5\%$ at peak-to-peak)
was found in the flat field frames of the filter
N-A-L659, the flat-fielded H$\alpha$+[\ion{N}{2}]
images have a large-scale non-uniform pattern over a CCD frame.
The cause of this non-uniformity has not yet been identified.
We fitted the pattern with
a cubic polynomial and subtracted the fitted function from the object frames
after the ordinary flat-fielding procedure.
Therefore, the internal accuracy of photometry in N-A-L659 frames should be
at worst $\sim \pm 5\%$, which is accurate enough for this study.

  The CCD was saturated at and around the nucleus of the galaxy in all
the object frames taken.  In the narrow-band frames, the region of
radius $r < 3$ arcsec centered on the nucleus was affected by
saturation.  In the broad-band frames, the saturation was more severe
and the whole central region ($r<15$ arcsec) could not be used for data
analysis (see Figs 1c and 1d).  Part of the inner spiral arms was
also saturated in the $R_c$-band frames (Fig. 1c).

  The flat-fielded frames in the same bands were aligned using
field stars and co-added to improve the signal-to-noise ratio.
For each of the co-added frames a quadratic curved surface
was fitted to the sky background
well outside the galaxy disk and subtracted from the image.
In order to make an uncontaminated H$\alpha$+[\ion{N}{2}] image,
the $R_c$-band frame was scaled
so that the average counts in the featureless part of the galaxy disk
equaled those in the H$\alpha$ frame; this scaled
$R_c$-band frame was then subtracted from the H$\alpha$+[\ion{N}{2}] frame.
When subtracting, PSF equalization was performed, following the method of
Alard and Lupton (1998).
The same procedure was applied to the [\ion{O}{3}] frame
and the $V$-band frame to extract a pure [\ion{O}{3}] emission-line image.

  Flux calibrations were carried out using our observations of
spectrophotometric standard stars.  As mentioned above, the observing
conditions were not photometric; fluxes of standard stars obtained at
similar elevations varied by up to $\sim 10\%$.  Moreover,
differences in the effective transmission wavelengths between the
narrow and broad observing bands reduce the photometric accuracy in
regions, such as the shoulder of the bulge, where large color
gradients may exist.  For the H$\alpha$+[\ion{N}{2}] ($\lambda_0 =
6590$ \AA) and the $R_c$ ($\lambda_0 \sim 6700$ \AA) bands, the central
wavelength difference is small enough that the color difference may be
ignored.  On the other hand, the color difference between the
[\ion{O}{3}] filter ($\lambda_0 = 5020$ \AA) and the $V$-band filter
($\lambda_0 \sim 5500$ \AA) is not negligible.  The flux difference
between $\lambda 5000$ \AA\ and $\lambda 5500$ \AA\ for Sb galaxies is
$\sim 5\%$, according to Coleman, Wu and Weedman (1980).

  We also examined the amount of emission-line flux that might have
fallen outside the filter band pass due to the redshift of the
object. We found that the loss of flux is $20\%$ at most, even if the
recession velocity of emission-line clouds is 300 km s$^{-1}$ over the
systemic velocity of NGC 4388.  Combining this uncertainty with
the flux calibration error due to the non-photometric conditions, we
estimate that the uncertainty in the flux of our narrow-band images is
typically $20\%$, but could be up to $50\%$ at worst.  Comparing our
data with the spectrophotometric observations of Petitjean and Durret
(1993), we found our flux at 35\arcsec\ away from the nucleus at
P.A. $= 55 \degr$ to be consistent with theirs to within $30\%$.

\section{RESULTS}

  The continuum-subtracted H$\alpha$+[\ion{N}{2}] and
[\ion{O}{3}]$\lambda 5007$ \AA\ images and the broad-band images of
NGC 4388 are shown in Figs. 1a -- 1d.  The H$\alpha$+[\ion{N}{2}]
image is the deepest emission-line image of NGC 4388 ever taken.  The
detection limits are as faint as $\sim 3\times10^{-18}$ erg
s$^{-1}$ cm$^{-2}$ arcsec$^{-2}$ for the H$\alpha$+[\ion{N}{2}] image (Fig. 1a), and
$\sim 1 \times 10^{-17}$ erg s$^{-1}$ cm$^{-2}$ arcsec$^{-2}$ for the [\ion{O}{3}]
image (Fig. 1b).

  A remarkable feature of these images is the large number of extended
emission-line filaments that are clearly seen in both the
H$\alpha$+[\ion{N}{2}] and [\ion{O}{3}] images. (Hereafter,
we call the region of these filaments the very extended
emission-line region, or VEELR.)  The VEELR extends very
far outside the galaxy disk ($r >1$\arcmin) between P.A. $\sim30\degr$
and P.A. $\sim65\degr$.  A sketch of the VEELR is shown in Fig. 2.  The
most distant cloud/filament seen in the H$\alpha$ image is at $\approx
430\arcsec$ ($\approx 35$ kpc) northeast of the nucleus.  This
distant cloud/filament is marginally seen also in the [\ion{O}{3}]
image (Fig. 1b).  A close-up view of the VEELR is shown in Fig. 3,
where major clouds/filaments are identified by numbers.  There seem to
be two streams in the VEELR; one starts at 90\arcsec\ from the nucleus
along P.A. $\approx 55^\circ$ and curves smoothly to the north at
P.A. $\approx 10\degr$, with a total length of up to 120\arcsec,
and the other extends
between P.A. $\approx 50^\circ$ and P.A. $\approx 65^\circ$ from
$120\arcsec$ from the nucleus out to more than 300\arcsec.  The
latter stream is further divided into two streams at $200\arcsec$ from
the nucleus, one of which is a smooth emission region extended along
P.A. $\approx 60^\circ$ (OF-1 in Fig. 3)
and the other is a faint string of clouds extended along
P.A. $\approx 0^\circ$ -- $-20^\circ$ (C22, C23, C24).  At $300\arcsec$
from the nucleus, there is a large complex of clouds (OF-2), and the stream
bends to the north, extending up to $430\arcsec$ away.  The most distant
clouds/filament complex (OF-5) is curved like a bow whose string is directed
toward the nucleus.  The luminosity in H$\alpha$+[\ion{N}{2}]
of a typical cloud/filament in the VEELR is $\sim 10^{37}$ erg
s$^{-1}$.

  The ionized gas mass of each cloud/filament is calculated using the relation
\begin{displaymath}
M_{\rm gas} = V \; N_{\rm e}({\rm rms}) \; f_{\rm v}^{1/2} \; m_{\rm H},
\end{displaymath}
where $V$ is the volume of the filament, $N_{\rm e}$(rms) is the rms
value of electron density, $f_{\rm v}$ is volume filling factor, and
$m_{\rm H}$ is the mass of a hydrogen atom.  $N_{\rm
e}$(rms) is derived from the following equation (Yoshida \& Ohtani
1993):
\begin{displaymath}
N_{\rm e}({\rm rms}) \; ({\rm cm}^{-3}) = 1.2 \times \left( \frac{V}{10^6 {\rm pc}^3} \right)^{-1/2} \;
\left( \frac{L(\rm H\alpha)}{10^{37} {\rm erg} \; {\rm s}^{-1}} \right)^{1/2}.
\end{displaymath}
We assumed Case B recombination and the electron temperature $T_{\rm e} = 10^4$ K
in this calcuration.
Although the volume filling factor $f_{\rm v}$ is not known for the
VEELR clouds, a typical value for such extranuclear emission-line gas
is $\sim 10^{-3} - 10^{-4}$ (Yoshida \& Ohtani 1993; Robinson et
al. 1994; Aoki et al. 1994).  Since the typical H$\alpha$ luminosity and
size of a cloud/filament in the VEELR are $\sim
10^{37}$ erg s$^{-1}$ and a few hundred pc, respectively, its
rms electron density is $\sim 0.3 - 0.4 $ cm$^{-3}$,
assuming a spherical geometry for the cloud.  Therefore, if the local
electron density of emission-line clouds is $N_{\rm e}({\rm local}) \sim
10$, we infer $f_{\rm v}$ $\sim 2\times10^{-3}$, because $N_{\rm
e}({\rm local})$ is related to $N_{\rm e}({\rm rms})$ and
$f_{\rm v}$ by the equation $N_{\rm e}({\rm rms}) = f_{\rm v}^{1/2} \; N_{\rm
e}({\rm local})$.  If this is the case, the mass of a typical cloud is
$\sim 10^3$ $M_\sun$.

  The sizes, distances from the nucleus, ionized gas masses, and
emission-line fluxes of the gas
clouds/filaments in the VEELR of NGC 4388 are listed in Table 2.
The volume of each cloud/filament was calcurated assuming a ellipsoid
whose major axis is the projected major axis of each cloud/filament for compact, 
bright ones (C1 -- C26).
We assumed a sheet-like morphology for the diffuse filaments (e.g filaments in
OF-1 -- OF-5) whose thickness along our line of sight is 100 pc.
The total mass of the ionized gas of the VEELR is $\sim 4 \times 10^6$
$f_{\rm v}^{1/2}$ $M_\sun$.  This is comparable to that of the NE
plume derived by VBCTM, $\sim 10^5$ $M_\sun$, if $f_{\rm v} \sim 1
\times 10^{-3}$.

  The morphology of the bright part of the emission-line region in our
images (the disk \ion{H}{2} region, the NE plume, and the SW cone) is
similar to that found in previous studies (Pogge 1988; Corbin et
al. 1988, VBCTM), but more detailed structure is revealed.  Fig. 4 shows
a close-up of the NE plume. (In order for the
spatial structure of the emission-line region to be clearly seen, PSF
equalization, which degrades the spatial resolution of the image, was
not performed in making this figure.) The NE plume consists of many
filaments whose widths are marginally resolved at our
resolution ($\approx 0\farcs8$) and are about 80 -- 100 pc ($\approx
1^{\prime\prime}$).

  A map of the emission-line intensity ratio
$I$([\ion{O}{3}])$/I$(H$\alpha$+[\ion{N}{2}]) is shown in Fig. 5.  This
map is very useful for investigating the excitation structure of the
outer emission-line region, although the detailed structure of the
gas excitation distribution near the nucleus is totally lost in this
figure, because of severe saturation of the broad-band images.
Nevertheless, a distinct cone-like structure of high excitation gas is
clearly seen to the south of the nucleus.  This ``ionization cone''
was first found by Pogge (1988) and its structure has been
investigated in several studies (Pogge 1988; Falcke et al. 1998,
VBCTM).  The north side of the ionization cone is traced by the NE plume
and a high-ionization cloud located at $10\arcsec$ -- $15\arcsec$ 
northwest of the
nucleus.  We note that filaments of the inner VEELR ($\lesssim
10$ kpc; C1 -- C8) also show high ionization,
$I$([\ion{O}{3}])$/I$(H$\alpha$)$\gtrsim 1$, suggesting that these
filaments are also excited by the nuclear ionizing radiation.
Interestingly, the filaments at about 11 kpc from the nucleus show
an internal excitation gradient; the $I$([\ion{O}{3}])$/I$(H$\alpha$) ratio
smoothly decreases with distance from the nucleus in one filament
(C2, C3, C4) (Fig. 5).  In addition, we note that a bright filament at
P.A.$=70^\circ$, 11 kpc away from the nucleus (C25) shows quite low
excitation ($I$([\ion{O}{3}])$/I$(H$\alpha$)$\sim 0.1$), although its
projected distance from the nucleus is the same as those of clouds
C9 -- C15.  There seems to be a sudden change in the excitation state of
the gas at around the line of P.A.$= 65^{\circ}$, which coincides
with the extrapolation of the northern edge line of the SW cone.
These findings suggest that the nuclear ionization cone excites most
of the VEELR gas.  The excitation mechanism of the VEELR will be
discussed further in the next section.

\section{DISCUSSION}

\subsection{The excitation mechanism of the very extended emission-line 
region of NGC~4388}

  Figure 6 shows the radial variation of the emission-line intensity
ratio, $I$([O III])$/I$(H$\beta$).  In this figure the H$\beta$ flux in
the VEELR is derived from the H$\alpha$+[\ion{N}{2}] flux by assuming
that (1) the gas is in Case B recombination with an electron 
temperature of $10^4$ K ($I$(H$\alpha$)$/I$(H$\beta$)
= 2.86; Osterbrock 1989), (2) the $I$([\ion{N}{2}])$/I$(H$\alpha$)
emission-line intensity ratio is 0.2, and (3) there is no dust
extinction.  Since most of the VEELR is located well outside the
galactic disk of NGC 4388, the above assumptions are adequate to
estimate the H$\beta$ flux.  The $I$([O III])$/I$(H$\beta$) ratio
decreases with distance from the nucleus.

  Although it is impossible to estimate the excitation mechanism of
the ionized gas uniquely using only the $I$([O III])$/I$(H$\beta$)
intensity ratio, the monotonic decrease of this ratio with radius
suggests that the same mechanism is responsible for exciting the gas in
these regions. The NE plume has a similar emission-line spectrum to
that of the nucleus; the spectrum is consistent with power-law
photoionization models with an ionization parameter $U\sim 10^{-3}$
(Pogge 1988; Petitjean \& Durret 1993).  Since the distance from the
nucleus of the VEELR is 3 -- 4 times that of the NE plume, the number
of ionizing photons from the nucleus that reach the VEELR is smaller by
1/9 -- 1/16 times.  Thus, assuming that the gas density decreases
with $r^{-1}$, one finds that $U$ for the VEELR is a few times
$10^{-4}$.  Power-law photoionization models predict that $I$([O
III])$/I$(H$\beta$) $\approx$ 1 -- 3 for such values of $U$.  This is
consistent with the observed values of the emission-line intensity ratio.

In the case of nuclear power-law photoionization,
the number of nuclear ionizing photons necessary to ionize the VEELR
is estimated by the following relation:
\begin{displaymath}
Q_{\rm u} = 4 \pi R^2 c N_{\rm e} U,
\end{displaymath}
where $R$ is the distance of an VEELR cloud from the nucleus, $c$ is 
the speed of light, $N_{\rm e}$ and $U$ are the electron density and
the ionization parameter of the cloud, respectively.
Assuming that $N_{\rm e} \sim 10$ cm$^{-3}$ and $U \sim 10^{-4}$,
one finds that $Q_{\rm u}$ is a several times $10^{53}$ photons 
s$^{-1}$ for a cloud at a distance of $\sim 10$ kpc from the nucleus.
The value of $Q_{\rm u}$ derived above is a typical one of nearby
type 1 Seyfert galaxies (e.g. Padovani and Rafanelli 1988;
Yoshida and Ohtani 1993; Aoki et al. 1996). 
Taking into account the anisotropic nature of the nuclear radiation,
we conclude that the nuclear power-law continuum of NGC~4388 is
capable of ionizing most of the VEELR gas.

  The observed values of $I$([O III])$/I$(H$\beta$) in the VEELR are,
on the other hand, also consistent with expectations for other excitation
mechanisms, such as OB star ionization (\ion{H}{2} regions) or shock
heating.  They agree with the predictions of models of H II regions
with moderate metal abundances ([O/H] $\sim$ [O/H]$_\sun$) (McCall,
Rybski \& Shields 1985).  However, there is no signature of ionizing
star clusters near the VEELR.  If the VEELR is ionized by UV radiation
emanating from the disk star-forming complex, a total of $\sim 2\times
10^{2}$ O5 stars are needed to excite a typical H$\alpha$ cloud in the
VEELR (L(H$\alpha$)$\sim 10^{37}$ erg s$^{-1}$, the mean distance from
the disk $\approx 10$ kpc, and typical size of a cloud is 300 pc).  On
the other hand, the total number of ionizing stars estimated from the
H$\alpha$ luminosity of the disk \ion{H}{2} complex is $\sim
3\times10^{2}$.  These two numbers are consistent with each
other.  However, the actual number of disk-ionizing stars needed to
ionize the VEELR would be much larger than this estimate, because a
significant part of the ionizing radiation is exhausted in ionizing
the surrounding dense gas from which the stars were born.  It is
unlikely that such a large star-forming complex resides in the disk of
NGC 4388.

  Shock heating is another candidate excitation mechanism for the VEELR.
Shocks whose velocities are $\sim 100$ km s$^{-1}$ can explain the observed
$I$([\ion{O}{3}])$/I$(H$\beta$)
ratios of the VEELR (Schull \& McKee 1979; Binette, Dopita \& Tuohy 1985;
Dopita \& Sutherland 1995).
The kinematics and emission-line intensity ratios of other species, such as
$I$([\ion{N}{2}])$/I$(H$\alpha$) or $I$([\ion{S}{2}])$/I$(H$\alpha$), of the
VEELR gas are needed to estimate the contribution from shock heating.
Nevertheless, we consider that shock heating is not the main source of
the excitation of the VEELR, because
the shock heating model cannot explain (1) the smooth radial decrease of
the $I$([\ion{O}{3}])$/I$(H$\beta$) ratio
or (2) the sudden change of the $I$([\ion{O}{3}])$/I$(H$\alpha$) ratio
between the inside and outside of the ionization cone.

  Therefore, it is likely that most of the VEELR is ionized and excited
primarily by the nuclear ionizing radiation, although detailed spectroscopic
study is needed to determine the excitation source precisely.  (Note that the
excitation mechanism of the low-ionization filaments outside the
ionization cone (C25 and C26) is not identified.  The [\ion{O}{3}]
emission is significantly fainter than the H$\alpha$ in those
filaments, and is not detectable in our images.)

\subsection{Origin of the very extended emission-line region of NGC~4388}

  The VEELR is very large ($> 30$ kpc) and extends only on
the northeastern side of the galaxy.  Here, we discuss the origin of
the VEELR of NGC 4388.

\subsubsection{AGN wind}

  Gas outflow phenomena have often been detected around active galactic nuclei
(e.g., Mediavilla \& Arribas 1995; Aoki et al. 1996; Arribas, Mediavilla
\& Garcia-Lorenzo 1996; Veilleux et al. 1998).
VBCTM suggested that the most plausible origin of the inner extraplanar
ionized nebula of NGC 4388 is a bipolar outflow,
which was originally accelerated by the radio ejecta from the nucleus.

  Energetically, the radio jet could have accelerated the ionized gas
in the VEELR.  The total mass of the VEELR gas is of
order 10$^5$ $M_\sun$, and its radial velocities should be as fast
as those of the NE plume gas, i.e., 100 -- 200 km s$^{-1}$.  Thus, the
kinetic energy of the ionized gas of the VEELR is $\sim 10^{52}$ erg.
Although the turbulent energy of the VEELR is not known because of a
lack of detailed velocity information, it is likely to increase the
total kinetic energy of the VEELR by a factor of only two or three.
VBCTM argued that the radio jet of NGC 4388 could accelerate material
that had a similar level of kinetic energy.

  There are some problems in this scheme for the origin of
the VEELR of NGC 4388.  First, there is a large discrepancy between
the P.A. of the radio jet and that of the VEELR.  The radio ejecta of NGC
4388 have P.A. $\sim 5\degr$ (Hummel \& Saikia 1991; Irwin, Saikia \&
English 2000), whereas the VEELR is extended in P.A. $\sim 40\degr -
60\degr$.  VBCTM suggested that a slight discrepancy between the P.A. of
the NE plume and the radio jet can be explained by buoyancy of the jet
as a result of a density gradient perpendicular to the galaxy disk.  The P.A.
discrepancy between the VEELR and the jet is, however, much larger, and
buoyancy effects alone cannot explain it.  Secondly, there is the size
difference between the radio jet and the VEELR.  The VEELR is
very large, reaching out to 35 kpc from the
nucleus, so that its age is of order 10$^8$ yr.  On the
other hand, the radio jet has a length of only 1 kpc.  Therefore, the
VEELR is more than an order of magnitude older than the radio jet, even if
they have similar outflow velocities.

  Large precession coupled with intermittent activity, or dense gas
that blocks the radio jet, might solve these problems, but we
consider attributing such peculiar behavior to the jet as highly
speculative.  We conclude that a nuclear outflow powered by the
radio jet is not a plausible origin of the VEELR gas.

\subsubsection{Debris of ancient starburst superwind}

  Strong starburst activity in the nuclear regions of a galaxy causes a
powerful outflow from the galaxy (Heckman, Armus \& Miley 1990).  Such
``superwinds'', induced by the collective effect of supernovae and
the stellar winds of OB stars in a starburst region, are sometimes
strong enough to blow the interstellar gas in galaxies into
intergalactic space.  For example, the superwind of the well-studied
starburst galaxy M82 has a mechanical luminosity of $\sim 2.5 \times
10^{42}$ erg s$^{-1}$ and a momentum flux of $\sim 2 \times 10^{34}$
dynes (Lehnert, Heckman \& Weaver 1999), and optical filaments
associated with the wind extend well outside ($> 3$ kpc from)
the galaxy disk (Bland \& Tully, 1988; Shopbell \& Bland-Hawthorn
1998).  Recently, an H$\alpha$-emitting region, which may be associated
with the superwind, was discovered about 12 kpc to the north of M82
(Devine \& Bally 1999; Lehnert et al. 1999).  The twin bubble of
H$\alpha$-emitting gas of Arp 220 has a size of 14 kpc (Armus, Heckman
\& Miley 1989; Ohyama et al. 2001).  The size of the VEELR of NGC 4388 is
comparable to, but somewhat larger than, those of the superwinds
associated with these powerful starburst galaxies.

  In fact, strong star-forming activity is occurring in the disk of NGC
4388, which is traced by all existing H$\alpha$ images (Pogge 1988;
VBCTM; this study).  The total H$\alpha$ luminosity of the NGC 4388
disk is $\approx 10^{40}$ erg s$^{-1}$, corresponding to a
star-formation rate of $10 M_{\sun}$ yr$^{-1}$, which is comparable
to that of the prototypical starburst galaxy M82.  Thus, it would be
natural to consider that NGC 4388 generates a huge outflow due to its
strong star-forming activity.

  The highly asymmetrical distribution of the VEELR gas is, however,
problematic for this superwind scenario.  A superwind that is
powerful enough to blow the interstellar gas into intergalactic
space should blow on both sides of the disk, because the energy
injected by the starburst is so high that the hot gas produced by the
starburst should expand well beyond the disk scale height
(e.g., Chevalier \& Clegg 1985; Tomisaka and Ikeuchi 1988; Strickland
and Stevens 2000).  However, the VEELR is distributed only in the
northeast quadrant of NGC 4388.  Even if we consider the
fact that the gas in the region is excited by the asymmetric nuclear
radiation, so that the shape of the line emitting gas traces the path
of the radiation, it is difficult to explain the total lack of any
extension in the southwestern direction using the superwind model.

\subsubsection{Ram pressure stripped gas}

  NGC 4388 is located close to the center of the Virgo cluster.  It
has been suggested that it is currently at the bottom of the cluster
potential because of its extremely high peculiar velocity, which is
over 1000 km s$^{-1}$ relative to the systemic velocity of the cluster
(Yasuda et al. 1997).  Together with the \ion{H}{1} gas deficiency and
the sudden truncation of the \ion{H}{1} gas within the galaxy disk,
this has led many authors to suggest that the interstellar gas of NGC
4388 is affected by strong ram pressure due to a high-speed
interaction with the intracluster medium (ICM), and that the outer
part of the interstellar gas is stripped from the galaxy by the ram
pressure (Cayatte et al. 1994).

  Petitjean and Durret (1993) suggested that the extraplanar gas
extending to $50\arcsec$ away from the nucleus of NGC 4388 (the NE
plume) could be this ram-pressure-stripped gas.  Here, we examine whether
the ram-pressure-stripping model can be applied to the VEELR.

  Ram-pressure-stripped gas is blown away from a galaxy preferentially
in the opposite direction to the galaxy's motion relative to the ICM
(Abadi, Moor \& Bower 1999).  NGC 4388 is an edge-on galaxy, whose
inclination angle is $78\degr$; its north side is the near side
(Pogge 1988; Ayani \& Iye 1989; VBCTM).  
The VEELR is extended to the northeast of the galaxy.  Hence, if the VEELR
is the ram-pressure-stripped gas due to the interaction between the
galaxy disk and the ICM, NGC 4388 must have quite a large transverse
velocity southwestwards in order to blow the interstellar gas to
such a high galactic latitude.  Considering the large radial
component of the velocity relative to the ICM ($\sim 1400$ km
s$^{-1}$), one finds that NGC 4388 must have a transverse velocity of
$\sim 1500$ km s$^{-1}$.  If this is the case, the total collision
velocity of NGC 4388 to the ICM is $\sim 2000$ km s$^{-1}$ and its
collision angle is approximately 30--40\degr.  Numerical simulations
of ram-pressure stripping made by Abadi et al. (1999) suggest that
when the collision velocity and the angle are 2000 km s$^{-1}$ and
45\degr, respectively, ram-pressure-stripped gas is extended up to
10 -- 15 kpc from the galaxy disk after 10$^8$ yr has passed since the
collision between the galaxy and the ICM (see Figures 3 and 4 of Abadi et
al. 1999).  Furthermore, the truncation radius of \ion{H}{1} gas and the
surface density of the disk of NGC 4388 are consistent with their
45\degr\ collision model (Abadi et al. 1999).  Therefore, the ram-
pressure-stripping hypothesis for the origin of the VEELR gas is a
promising one, if NGC 4388 indeed has such very large transverse
velocity.

  However, VBCTM have argued that the transverse velocity of NGC 4388 is
rather small with respect to its radial velocity, so that the galaxy
is falling into the cluster core almost along our line of sight.  In
other words, NGC 4388 is colliding with the ICM nearly edge-on.  If
this were the case, ram-pressure-stripped gas would be blown along the
galaxy disk (Abadi et al. 1999), and the VEELR should be a projection
of a flow along the disk plane.  Since the projected distance of the
far-end of the VEELR from the galaxy disk is 10 kpc, the real distance
would be as much as $\approx 10 \times \cos^{-1} i \sim 50$kpc (due to
the inclination of the disk ${i=78\degr}$).  The morphology of the
VEELR strongly suggests that the gas extends perpendicular to the
galactic disk, and that such a flat elongation is not
likely. Moreover, such extremely elongated filaments in the plane of
the galaxy disk could not be illuminated by the nuclear ionization
cone indicated by the circumnuclear ionization structure.
Consequently, if the transverse velocity of NGC 4388 is small, it is
hard to believe that the VEELR gas is the interstellar gas stripped by
ICM ram pressure.

  Therefore, knowledge of the value of the transverse velocity of NGC
4388 is the key to determining whether the ram-pressure-
stripping hypothesis is plausible.  It is, however, impossible to
measure the transverse velocity directly.  Hence, we cannot conclude
that ram-pressure stripping is responsible for the VEELR.  To
investigate this subject further, it would be useful to know the
velocity field of the VEELR, which could be measured by deep
spectroscopy.

  We note that the outer edge of the VEELR may be strongly affected by
interaction with the ICM.  According to VBCTM, a large Mach cone with
an opening angle of $\approx 80^{\circ}$ is formed around NGC 4388 by
the interaction with the Virgo cluster ICM.  The apex of the cone is
at the leading edge of the \ion{H}{1} disk of the galaxy.  If this is
the case, the height of the contact discontinuity of the bow shock is
$= R_{\rm HI} \times \tan 40\degr \approx 8 - 9$ kpc above the galaxy
plane.  This is well inside the extension of the VEELR.  Thus, the outer
part of the VEELR may be severely affected by the shock.  The peculiar
morphology of the VEELR, particularly its apparent bending, may be due
to this interaction with the ICM.

\subsubsection{Tidal debris from galaxy-galaxy interaction}

  Pogge (1988) suggested that the NE plume of NGC 4388 might be tidal
debris from a past galaxy-galaxy interaction, like the Magellanic
stream.  VBCTM rejected the tidal debris scenario, because they did
not detect any distorted kinematic signatures in the disk velocity
field, and because it does not explain the correlation between the
[\ion{O}{3}] surface brightness and the kinematics of the NE plume
clouds (blue-shifted clouds tend to have high surface brightness).
Considering the large size and stream-like morphology of the VEELR,
however, we consider that a tidal debris model is a good candidate
to explain the origin of the VEELR gas.

  NGC 4388 lies near the core of the Virgo cluster (Yasuda et al. 1997),
where galaxy-galaxy encounters may be comparatively frequent.
Phillips and Malin (1982) discussed the morphological anomaly of the
outskirts of the disk of NGC 4388.  They commented on a faint envelope
surrounding the western edge of the galaxy.  We have also found that the
disk of NGC 4388 is moderately asymmetrical at a low surface
brightness level in our broad-band images, and that a faint hump exists
near the southwest edge ($\sim 300\arcsec$ away from the
nucleus) of the galaxy (see Figs. 1c and 1d).  This hump is seen in
both the $R_c$- and $V$-band images, and is not found in the pure
emission-line image, so that it should be mainly composed of stars.
In addition, the bulge of NGC 4388 is boxy (Veilleux et al. 1999b).
The features described above are clearly seen in Fig. 7, which shows the
isophotes of NGC 4388 made from our $R_c$-band images (Fig. 1c).
It has gradually come to be recognized that galaxies with
boxy bulges have experienced minor merging in the past (Mihos et
al. 1995; Walker, Mihos \& Hernquist 1996).

  The numerical simulations of minor merging carried out by Walker et
al. (1996) indicate that a minor merging excites a strong bar mode ($m
= 2$) instability in 1 Gyr after the merging and thickens the disk of
the primary galaxy significantly. The edge-on view of the disk and
satellite particle 1 Gyr after merging in the simulations of Walker
et al. (1996) strikingly resembles our R-band image of NGC 4388. The
thick, asymmetric morphology of the disk with faint humps extending
perpendicular to the disk plane at its edge, seen in their
simulations, are also present in our image, indicating that NGC 4388
may be a minor merger remnant seen $\sim 1$ Gyr after the beginning of
the merger event.  If this was the case, and the satellite was a
gas-rich dwarf galaxy, a long tidal tail of gas from the satellite
would be left along the path of the encounter.  In fact, such large
tidal tails of gas have been found around some mergers in \ion{H}{1}
observations (Arp 219, Smith 1994; NGC 7252, Hibbard et al. 1994; Arp
299, Hibbard and Yun 1999).  The morphological similarity between the
results of numerical simulations (Hernquist \& Mihos 1995; Walker et
al.  1996) and our R-band image of NGC 4388, together with the
stream-like morphology of the VEELR, lead us to a conclusion that NGC
4388 is a minor merger remnant and that the VEELR consists of dense gas clouds
in the tidal tails (tidal debris) illuminated by nuclear power-law
radiation.

  A minor merger also concentrates interstellar gas in the nucleus of
the primary galaxy, and may drive nuclear activity.  Taniguchi (1999)
has argued that minor mergers are one of the most plausible candidate
processes for driving Seyfert activity.  He proposed that Seyfert 2
galaxies are activated by a minor merger with a highly inclined orbit.
If this is the case for NGC 4388, a past minor merger may not only
have formed the tidal tail in which the VEELR gas clumps are embedded,
but may also have induced its nuclear activity.

\section{Summary}

  We have found an very extended emission-line region (VEELR) that
extends up to $\sim 35$ kpc from the type 2 Seyfert 2 galaxy NGC 4388.
This region consists of many faint gas clouds/filaments and
extends northeastwards from the galaxy.  The map of emission-line
intensity ratio, $I$([\ion{O}{3}])$/I$(H$\alpha$), suggests that the
clouds in the inner ($r<12$ kpc) region of the VEELR are excited by
nuclear ionizing radiation.  Although the excitation mechanism of the
clouds in the outer ($r>12$ kpc) region is unclear, those clouds might
also be excited primarily by the nuclear radiation, since they are
within the nuclear ionization cone.  We discussed the origin of
this ionized gas, and concluded that it is most plausible that the
VEELR gas is tidal debris due to a past interaction with a gas-rich
dwarf galaxy.  If NGC 4388 has a large transverse velocity, of the
order 10$^3$ km s$^{-1}$ toward the south of the galaxy, it is
also possible that the VEELR was formed by ram-pressure
stripping of the disk gas due to a high-speed collision between the
galaxy disk and the hot ICM.

  In any case, since the VEELR lies at a very high latitude with respect
to the disk plane, it is certain that the VEELR contains valuable
information about the halo gas of NGC 4388.  Deep spectroscopy of the
VEELR will not only tell us the nature of the VEELR gas, giving us more
reliable information on its kinematics, excitation mechanism, and
metal abundance, but will also allow us to consider the origin of the
halo gas of NGC 4388.  In this context, we emphasize that very
extended faint ionized gas that is illuminated by powerful radiation,
such as AGN radiation, may provide us with a unique opportunity to
investigate the physical state of halo gas around galaxies.
Wide-field, deep, narrow-band imaging and deep spectroscopy of the
halos of galaxies exhibiting AGN activity will be useful in such
studies.

\acknowledgments

  We are grateful to the staff of the Subaru telescope for their kind
help with the observations.
We thank Y. Taniguchi and T. Nagao for useful discussions.
We also thank anonymous refree for her/his comments on the manuscript. 
In addition, M. Y. thanks the staff of Okayama Astrophysical Observatory
for their encouragement during the course of this study.
This work was done using the facilities at the Astronomical Data
Analysis Center, National Astronomical Observatory of Japan.
This research has made use of NASA's Astrophysics Data System Abstract
Service.


\newpage


\figcaption[]{
\label{Figure 1}
Narrow-band and broad-band images of NGC 4388.  In each image, the
brightest regions are blacked out in order to enhance the
contrast of faint features. 
North is up and east is to the left. The white, star-like features around the
nucleus are artifacts caused by severe saturation of the CCD.  a) A
continuum subtracted H$\alpha$+[\ion{N}{2}] image.  Faint, highly
extended emission-line gas (the very extended emission-line
region, or VEELR) is clearly seen in the northeastern quadrant of the
galaxy halo.  b) A continuum subtracted [\ion{O}{3}]$\lambda5007$ \AA
image.  Bright filaments in the inner VEELR can clearly be seen in this
image, whereas the outer VEELR is very faint and hardly visible.  c) An
$R_c$-band image.  Note that the continuum light distribution of NGC
4388 is asymmetric at a faint surface brightness level.  A faint
``hump'' and ``tail'' are seen at the southwest edge of the disk and
at the west edge of the disk, respectively.  d) A $V$-band image.}

\figcaption[]{
\label{Figure 2}
A sketch of the distribution of the H$\alpha$-emitting gas around NGC 4388.}

\figcaption[]{
\label{Figure 3}
Close-up H$\alpha$+[\ion{N}{2}] images of the VEELR. The detailed
structure of the inner VEELR can clearly be seen.  The physical
parameters of the filaments/clouds numbered in this figure are
presented in Table 2.}

\figcaption[]{
\label{Figure 4}
A close-up [\ion{O}{3}] image of the NE plume.
The FWHM of the PSF of this image $\approx 0\farcs8$, corresponding to
65 pc at the object.
Many filaments/clouds are marginally resolved with this spatial resolution.}

\figcaption[]{
\label{Figure 5}
A map of the $I$([\ion{O}{3}])$/I$(H$\alpha$) emission-line intensity ratio of
NGC 4388.  The nuclear ionization cone is clearly revealed in the southwest
of the nucleus, and the disk star-forming regions can be seen
as low-ionization complexes.}

\figcaption[]{
\label{Figure 6}
The radial dependence of the $I$([\ion{O}{3}])$/I$(H$\beta$)
ratio of the VEELR.}

\figcaption[]{
\label{Figure 7}
An isophote map of the $R_c$ band image of NGC 4388.  The shape of the
bulge is boxy rather than spherical.  The outer isophotal contour is
moderately asymmetrical with respect to the nucleus.  A faint hump and
tail are seen at the west-southwest edge and at the western edge of
the disk, respectively.}

\newpage

\begin{deluxetable}{lcccc}
  \tablewidth{0pt}
  \tablecaption{Journal of the Observations}
  \tablehead{
    \colhead{Filter ID} & \colhead{$\lambda_0$(\AA)} &
    \colhead{$\Delta\lambda$(\AA)} & \colhead{Date} &
    \colhead{Exposure Time} 
  }
  \startdata
    N\_A\_L659 & 6600 & 100   & 2001 April 26 & 1200 sec $\times$ 8 \\
    N\_A\_L503 & 5020 & 100   & 2001 April 24 & 1200 sec $\times$ 3 \\
    R$_c$-band & 6700 & 1000 & 2001 March 24 & 240  sec $\times$ 3 \\
    V-band    & 5500 & 800  & 2001 March 24 & 300 sec $\times$ 3 \\
  \enddata
\end{deluxetable}

\newpage

\begin{deluxetable}{lrrrrcrrr}
  \tabletypesize{\footnotesize}
  \tablewidth{0pt}
  \footnotesize
  \tablecaption{Physical Parameters of the VEELR Gas Clouds}
  \tablehead{
    \colhead{Cloud ID} & 
    \colhead{$f_{\rm H\alpha+[N II]}$ \tablenotemark{a}} &
    \colhead{$f_{\rm [O III]}$ \tablenotemark{b}} &
    \colhead{$L_{\rm H\alpha+[N II]}$ \tablenotemark{c}} &
    \colhead{$L_{\rm [O III]}$ \tablenotemark{d}} & 
    \colhead{size\tablenotemark{e}} &
    \colhead{distance\tablenotemark{f}} & 
    \colhead{$N_{\rm e}$(rms) \tablenotemark{g}} &
    \colhead{$M_{\rm gas}$ $f_{\rm V}^{-1/2}$ \tablenotemark{h}} 
  }
  \startdata
    C1  & 9.4 & 8.3 & 3.1 & 2.7 & 160$\times$320 & 7.6 & 0.32 & 3.2 \\
    C2  & 17.7  & 18.5 & 5.9 & 6.2 & 190$\times$310 & 8.4 & 0.37 & 5.2 \\
    C3  & 22.3  & 17.8 & 7.4 & 5.9 & 190$\times$360 & 8.8 & 0.39 & 6.2 \\
    C4  & 6.6 & 4.0 & 2.2 & 1.3 & 110$\times$240 & 9.1 & 0.44 & 1.6 \\
    C5  & 10.7  & 4.5 & 3.6 & 1.5 & 160$\times$490 & 10.0 & 0.28 & 4.2 \\
    C6  & 3.0 & 3.4 & 1.0 & 1.1 & 100$\times$240 & 7.9 & 0.35 & 1.0 \\
    C7  & 17.9  & 18.6 & 5.9 & 6.2 & 5.3\tablenotemark{i} & 8.5 & 0.40 & 4.8 \\
    C8  & 29.7  & 19.7 & 9.9 & 6.6 & 190$\times$530 & 9.7 & 0.37 & 8.8 \\
    C9  & 5.2  & 2.0 & 1.7 & 0.7 & 130$\times$280 & 10.4 & 0.32 & 1.8 \\
    C10  & 6.1  & 3.2 & 2.0 & 1.1 & 110$\times$240 & 10.2 & 0.20 & 1.6 \\
    C11  & 31.9  & 19.4 & 10.6 & 6.5 & 210$\times$340 & 11.1 & 0.44 & 7.9 \\
    C12  & 6.5  & 4.2 & 2.2 & 1.4 & 2.5\tablenotemark{i} & 10.9 & 0.35 & 2.0 \\
    C13  & 11.2  & 4.3 & 3.7 & 1.4 & 130$\times$210 & 10.5 & 0.54 & 2.3 \\
    C14  & 12.4  & 4.9 & 4.1 & 1.6 & 110$\times$400 & 11.4 & 0.47 & 2.9 \\
    C15  & 8.4  & 4.0 & 2.8 & 1.3 & 130$\times$230 & 11.9 & 0.45 & 2.0 \\
    C16  & 11.9  & 4.9 & 4.0 & 1.6 & 110$\times$340 & 13.2 & 0.50 & 2.6 \\
    C17  & 20.8  & 8.2 & 6.9 & 2.7 & 3.6\tablenotemark{i} & 12.5 & 0.53 & 4.3 \\
    C18  & 15.7  & 9.2 & 5.2 & 3.1 & 100$\times$630 & 13.8 & 0.49 & 3.5 \\
    C19  & 6.4 & 2.7 & 2.1 & 0.9 & 100$\times$230 & 14.8 & 0.52 & 1.3 \\
    C20  & 3.2  & 0.3 & 1.1 & 0.1 & 60$\times$190 & 14.9 & 0.60 & 0.6 \\
    C21  & 4.0  & \nodata & 1.3 & \nodata & 100$\times$230 & 14.0 & 0.42 & 1.1 \\
    C22  & 23.4  & 8.9 & 7.8 & 3.0 & 7.6\tablenotemark{i} & 15.6 & 0.39 & 6.6 \\
    C23  & 13.1  & \nodata & 4.4 & \nodata & 160$\times$890 & 15.5 & 0.23 & 6.3 \\
    C24  & 5.3  & \nodata & 1.8 & \nodata & 100$\times$190 & 15.5 & 0.51 & 1.1 \\
    C25  & 96.0  & 13.0 & 31.8 & 4.3 & 21\tablenotemark{i} & 11.8 & 0.55 & 22 \\
    C26  & 167.0  & 33.1 & 55.6 & 11.0 & 32\tablenotemark{i} & 10.1 & 0.5 & 36 \\
    OF-1  & 87.0  & \nodata & 28.9 & \nodata & 117\tablenotemark{i} & 18 & 0.2 & 49 \\
    OF-2  & 348.0  & \nodata & 116.0 & \nodata & 141\tablenotemark{i} & 24 & 0.34 & 110.0 \\
    OF-3  & 11.9  & \nodata & 4.0 & \nodata & 12.6\tablenotemark{i} & 22 & 0.2 & 6.1 \\
    OF-4  & 24.3  & \nodata & 8.1 & \nodata & 104\tablenotemark{i} & 31 & 0.1 & 25 \\
    OF-5  & 156.0  & \nodata & 52.0 & \nodata & 184\tablenotemark{i} & 35 & 0.2 & 85 \\
  \enddata

  \tablenotetext{a}{H$\alpha$+[\ion{N}{2}] flux in units of 10$^{-17}$ erg s$^{-1}$ cm$^{-2}$.}
  \tablenotetext{b}{[\ion{O}{3}]$\lambda 5007$\AA\  flux in units of 10$^{-17}$ erg s$^{-1}$ cm$^{-2}$.} 
  \tablenotetext{c}{H$\alpha$+[\ion{N}{2}] luminosity in units of 10$^{36}$ erg s$^{-1}$.
The distance of NGC~4388 is assumed to be 16.7 Mpc.}
  \tablenotetext{d}{[\ion{O}{3}]$\lambda 5007$\AA\  luminosity in units of 10$^{36}$ erg s$^{-1}$.}
  \tablenotetext{e}{cloud size: pc $\times$ pc.}
  \tablenotetext{f}{distance from the nucleus: kpc.}
  \tablenotetext{g}{rms electron density in units of cm$^{-3}$.}
  \tablenotetext{h}{ionized gas mass in units of $10^4 M_{\sun}$.}
  \tablenotetext{i}{cloud volume in  units of 10$^6$ pc$^3$.}
\end{deluxetable}

\newpage
\begin{figure}
\figurenum{2}
\plotone{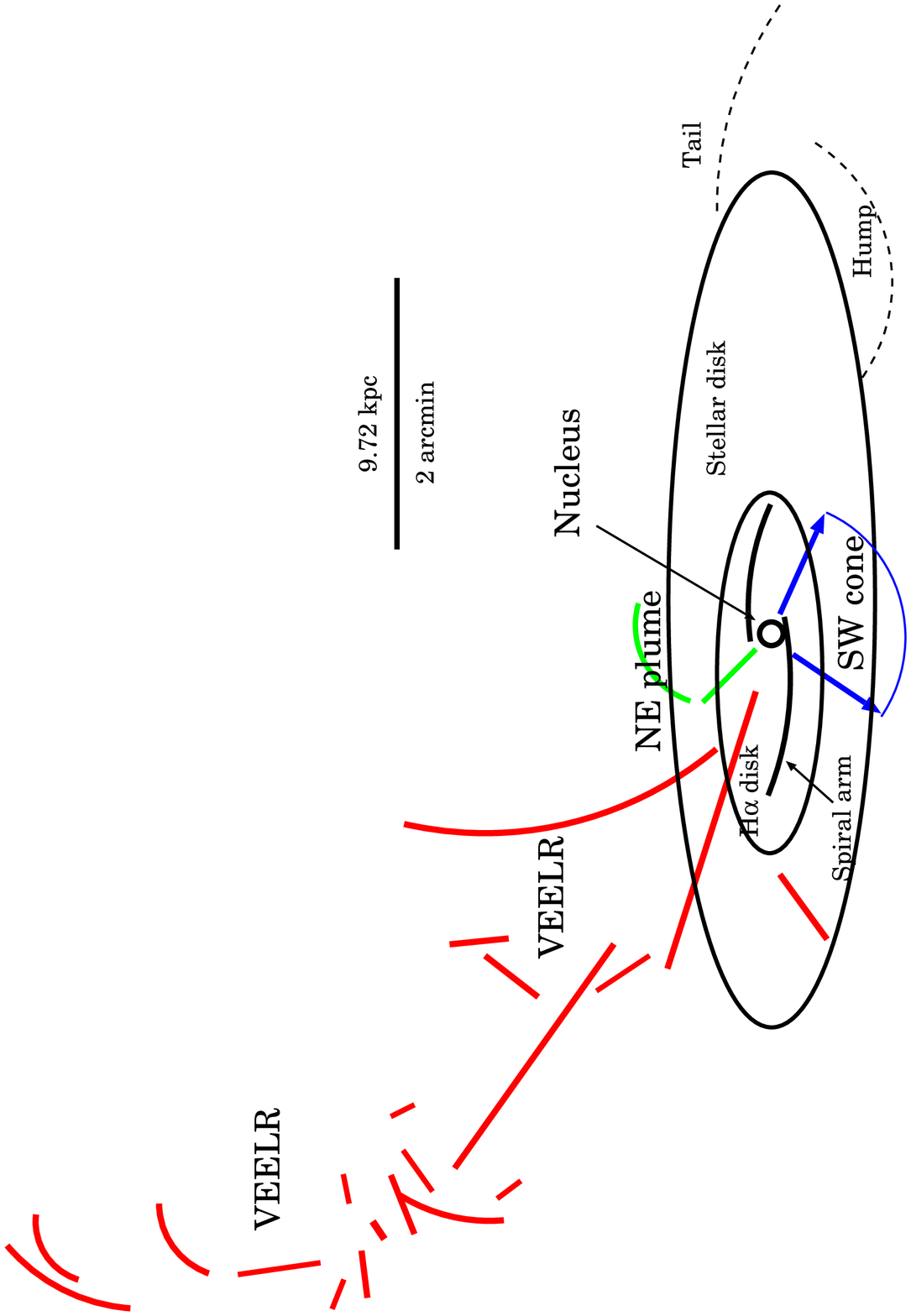}
\caption{Sketch of the VEELR.}
\end{figure}

\newpage
\begin{figure}
\figurenum{6}
\plotone{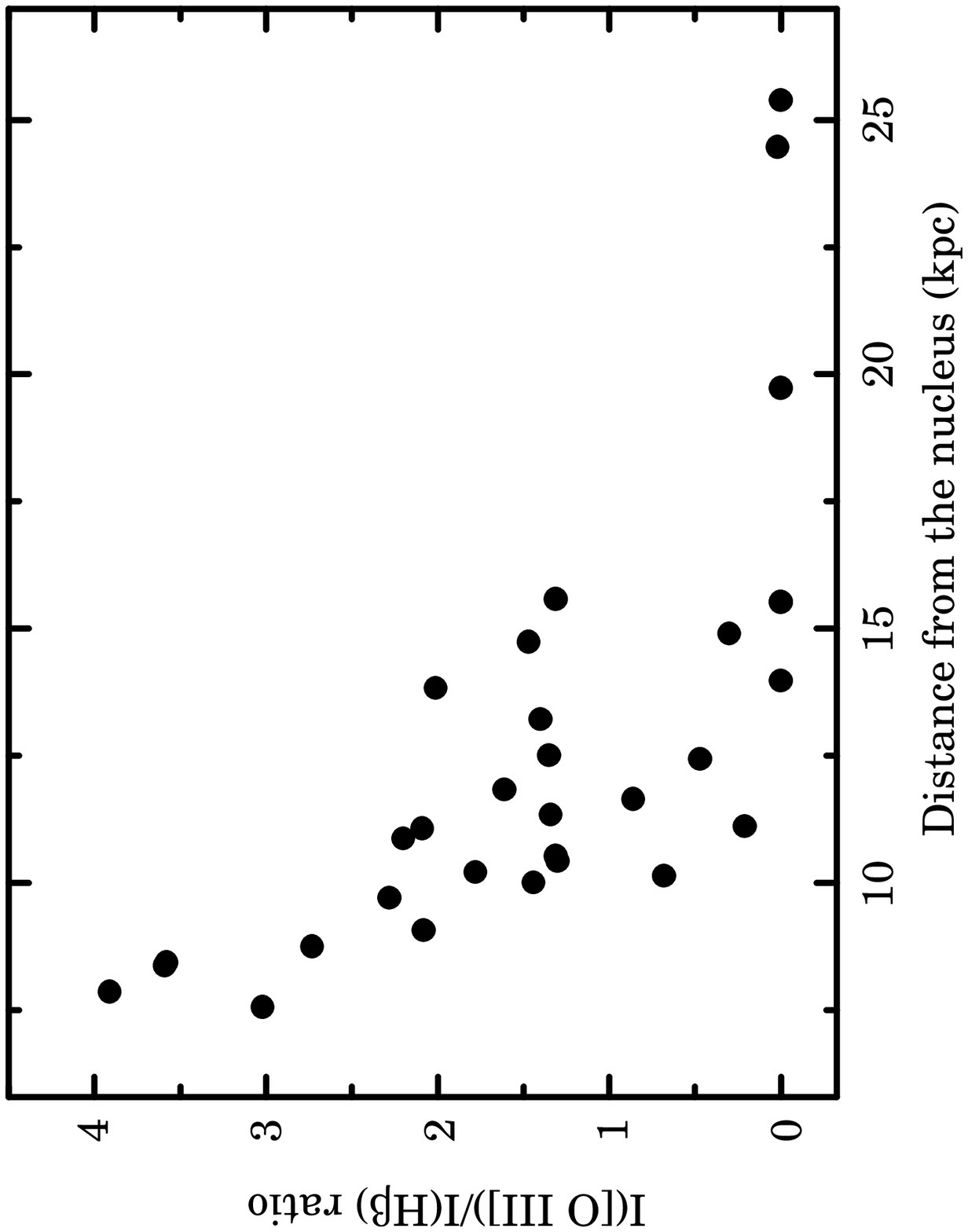}
\caption{The radial dependence of the I(ion{O}{3}])/I(H$\beta$) ratio.}
\end{figure}

\end{document}